# A SOFTWARE DECODER IMPLEMENTATION FOR H.266/VVC VIDEO CODING STANDARD

*Bin Zhu, Shan Liu, Yuan Liu, Yi Luo, Jing Ye, Haiyan Xu, Ying Huang, Hualong Jiao, Xiaozhong Xu, Xianguo Zhang, Chenchen Gu*

## ABSTRACT

Versatile Video Coding Standard (H.266/VVC) was completed by Joint Video Expert Team (JVET) of ITU-T and ISO/IEC, in July 2020. This new ITU recommendation/international standard is a successor to the well-known H.265/HEVC video coding standard with roughly doubled compression efficiency, but also at the cost of an increased computational complexity. The complexity of H.266/VVC decoder processing modules is studied in this paper. An optimized decoder implementation using SIMD instruction extensions and additional parallel processing including data and task level parallelism is presented, which can achieve real-time decoding of 4K 60fps VVC bitstreams on an x86 based CPU.

## 1. INTRODUCTION

The Versatile Video Coding Standard (H.266/VVC) [1] development was initiated by the Joint Video Experts Team of ITU-T VCEG and ISO/IEC MPEG in April 2018, and the standard was finalized in July 2020. H.266/VVC can provide up to 50% improvement in compression efficiency for the same subjective quality compared to its predecessor HEVC [2]. H.266/VVC is aiming at supporting a wide variety of video applications including ultra-high-definition video (e.g. 4K or 8K resolution), video with high dynamic range and wide color gamut, and video for immersive media applications such as 360° omnidirectional video, in addition to the applications that have commonly been addressed by prior video coding standards. H.266/VVC supports scalable coding in both spatial domain and temporal domain and provides screen content coding tools for the gaming and screen sharing applications. In order to reduce the bitrate of the inter coding, several coding tools with intensive decoder side computations are adopted in the H.266/VVC standard, such as decoder side motion vector refinement (DMVR), bi-direction optical flow (BDOF), and prediction refinement with optical flow (PROF). These coding tools introduce additional complexities for the decoder, thus the H.266/VVC reference test model decoder is 96% slower than the H.265/HEVC reference test model decoder on average [3].

While broadcast applications or low power devices usually require hardware decoder chips, there are still lots of people watching internet video on their computer where software decoding plays an important role. Besides this, hardware decoder chips normally come a few years later after a standard is finalized. Therefore, it is essential to have an efficient and optimized software decoder implementation to support all those new emerging applications. This paper demonstrates a highly optimized real-time H.266/VVC software decoder implementation and the remainder of this paper is structured as follows: Section 2 presents an overview of H.266/VVC decoder processing blocks and analyzes their complexities. Section 3 elaborates our software decoder design including SIMD optimizations, bit depth templatization and data/task level parallelism. Section 4 reports profiling results of our proposed decoder for 1080P HD/4K UHD test sequences with both 8bit and 10bit as input. Section 5 introduces a H.266/VVC video player application that we have developed. Finally, section 6 concludes the paper and also gives an outlook of our future work.

## 2. H.266/VVC DECODER OVERVIEW

A general H.266/VVC decoder structure and its main processing blocks are depicted in Fig. 1. Entropy decoder stage extracts binary codewords from the input bitstream and converts them to syntax elements including block partitioning structures, intra/inter prediction modes, quantized transforms coefficients (TCOEFFs), motion vectors (MVs), and reference frame indexes (RefIdxs). VVC uses context-adaptive binary arithmetic coding (CABAC) for entropy coding. It is similar to the CABAC scheme in HEVC but with several enhancements to improve its throughput speed and its compression performance.

Inverse quantization and inverse transform (IQ/IT) stage dequantizes and transforms frequency-domain TCOEFFs back to spatial domain residual blocks. In addition to DCT-II which has been employed in H.265/HEVC, H.266/VVC adopts a Multiple Transform Selection (MTS) scheme which includes some newly introduced transforms (DST-VII and DCT-VIII), and the transform matrices are quantized more accurately than the transform matrices in H.265/HEVC. H.266/VVC also introduces a low-frequency non-separable transform (LFNST) that is applied between inverse quantization and inverse primary transform. In LFNST, a 4x4 or 8x8 non-separable transform is applied according to the primary transform block size.

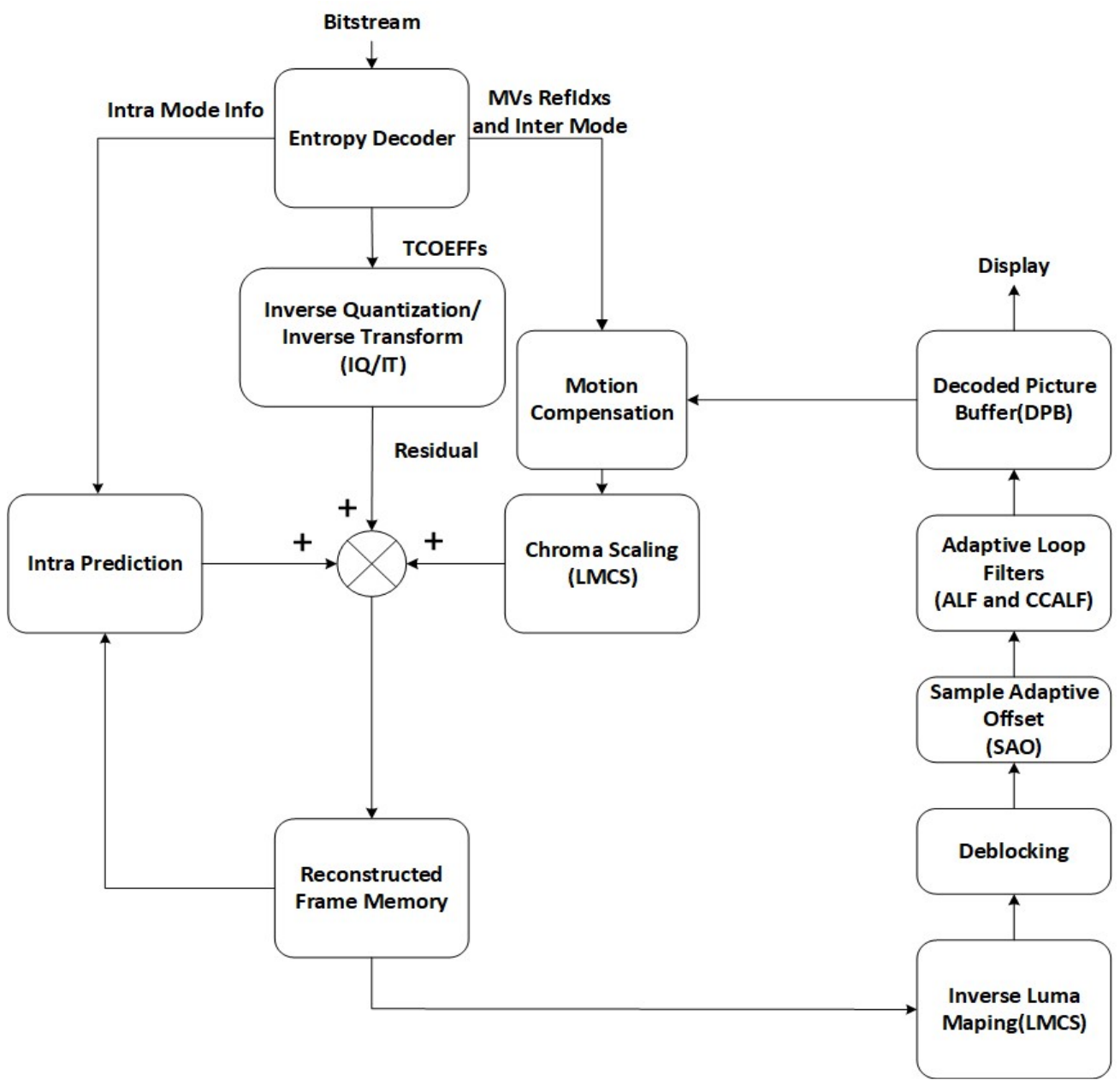

**Fig. 1** VVC Decoder Structure

Intra prediction stage computes intra prediction for a decoded block. H.266/VVC extends angular intra prediction modes to 65 from 33 as used in H.265/HEVC and several conventional angular intra prediction modes are adaptively replaced with wide-angle intra prediction modes for the non-square blocks. To reduce the cross-component redundancy, a cross-component linear model (CCLM) prediction mode is used in the H.265/VVC, for which the chroma samples are predicted based on the reconstructed luma samples of the same CU. If the decoder operates in intra prediction mode, intra predicted block is added to residual block to form the reconstructed block.

Inter prediction stage produces motion compensated (MC) inter prediction for a decoded block by referencing the decoded picture buffer (DPB) with MVs and RefIdxes. DPB contains previously decoded pictures which can be used for both short-term and long-term references. H.266/VVC uses 8-tap interpolation filter for luminance and 4-tap filter for chrominance samples in 1/16-pixel, 1/4-pixel, and 1/2-pixel MC. In H.266/VVC, motion vector prediction and coding have been improved by using technologies such as merge with MVD (MMVD), history based MV prediction (HMVP), symmetric MVD and adaptive MV resolutions (AMVR). To improve sample prediction accuracy, several new merge modes have been added, including combination of inter and intra prediction (CIIP), geometrical partitioning mode (GPM) and bi-prediction with CU-level weights (BCW) modes. In addition to the translational motion model used in H.265/HEVC, a subblock-based affine motion model can also be applied in H.266/VVC. The affine motion field of the block is described by motion information of two control-point (4-parameter) or three control-point motion vectors (6-parameter). If the decoder operates in inter prediction mode, inter predicted block is added to the residual block to form the reconstructed block.

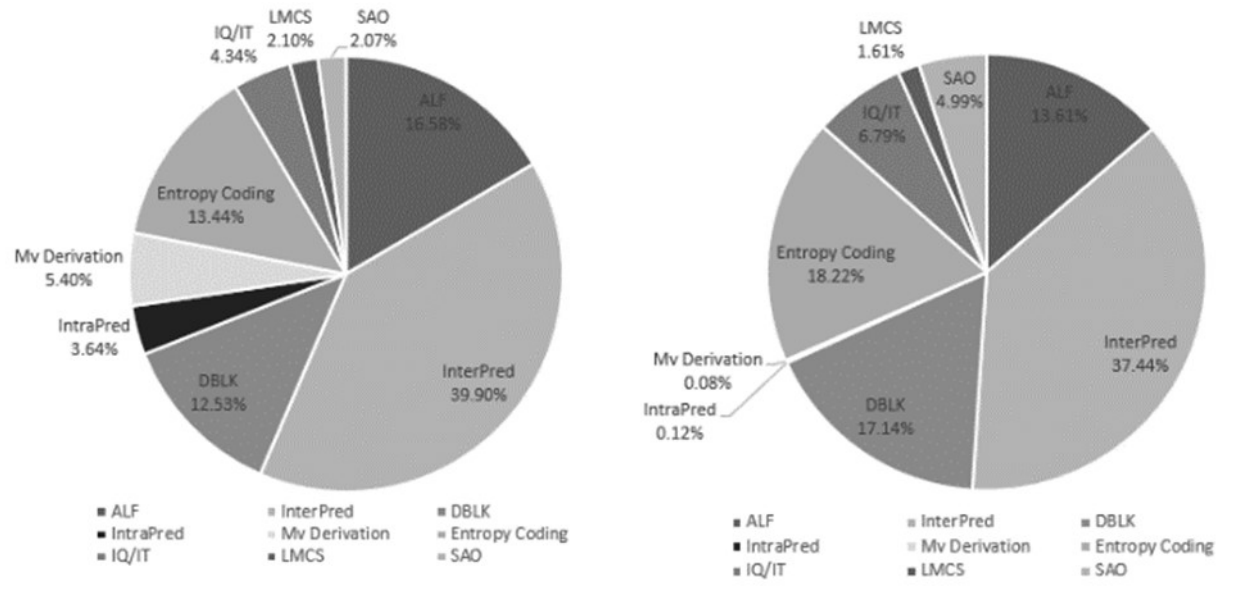

**Fig. 2** Breakdown of Decoding Time for Random Access Configuration at QP = 22 (left: our decoder, right: VTM)

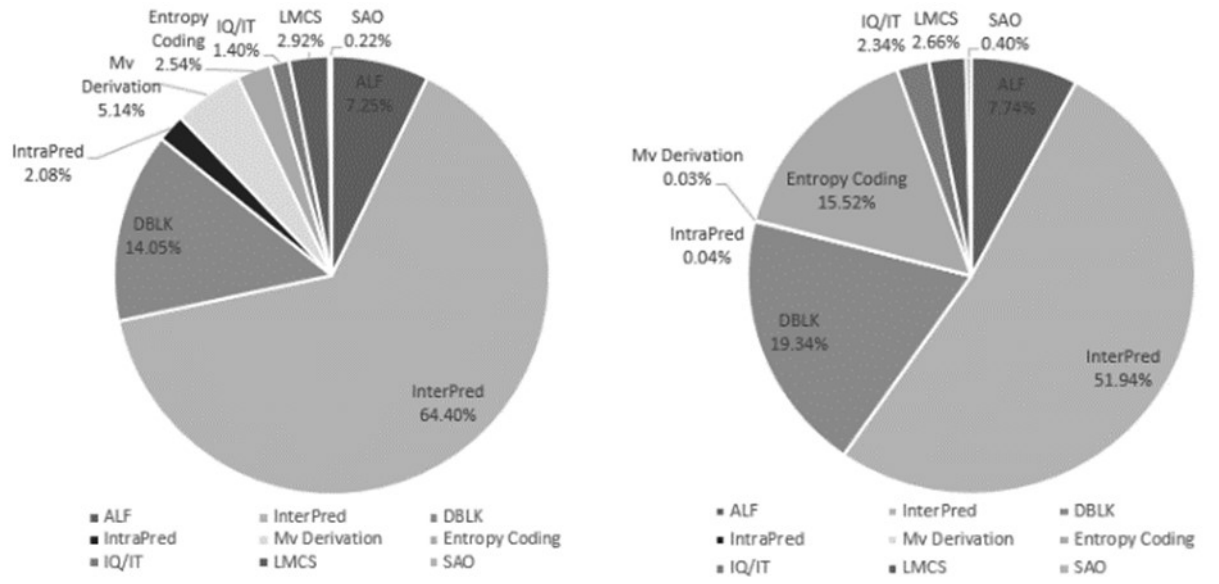

**Fig. 3** Breakdown of Decoding Time for Random Access Configuration at QP = 37 (left: our decoder, right: VTM)

Luma mapping with chroma scaling (LMCS) stage is a new processing block added in H.266/VVC before the loop filters. LMCS has two main components: 1) in-loop mapping of the luma component based on adaptive piecewise linear models; 2) luma dependent chroma residual scaling is applied for chroma components.

Loop filtering (LF) stage filters the reconstructed pictures. LF stage contains three in-loop filters: deblocking filter (DBLK), sample-adaptive offset (SAO) and adaptive loop filter (ALF). Deblocking filter removes the blocking artifacts at the block boundary, including coding unit (CU) boundaries, transform unit (TU) boundaries, and prediction subblock boundaries. For luma deblocking, the filter is applied on a 4x4 grid, which is different from H.265/HEVC. The SAO filtering in H.266/VVC is the same as that in H.265/HEVC. Adaptive loop filters consists of luma/chroma ALF with block-based filter adaption applied, and cross-component adaptive loop filter (CCALF) that uses luma sample values to refine each chroma component by applying an adaptive, linear filter to the luma channel and then using the output of this filtering operation for chroma refinement.

Comprehensive H.266/VVC tool tests were conducted by AHG13 of JVET with analysis results reported in [4]. Fig. 2 and Fig. 3 show a rough breakdown of the runtime spent on each processing block with our decoder and H.266/VVC test model (VTM) decoder running in single-thread mode, the test bitstreams are encoded from one of the test sequences used in H.266/VVC common test conditions (CTC) [5] with random-access (RA) configuration and quantization parameter (QP) value set to 22 and 37 separately. Note the optimization level is not homogeneous across all the processing blocks. The results show that inter prediction is the most time-consuming block in the whole decoding pipeline and LF filters also take a significant portion of the decoding time given their complexities. At low QP range, more time would be spent on entropy decoding.

## 3. H.266/VVC DECODER OPTIMIZATIONS

Our VVC software decoder has been implemented from scratch and optimized with SIMD intrinsic for AVX2 extensions. In order to efficiently support different input bit depth, we have designed the decoder with C++ class and function template [6]. Multithreading has been performed using the C++ 11 thread libraries to make it portable across multiple operating systems.

### 3.1 SIMD Optimization

Similar to SIMD optimization in H.265/HEVC decoding [7], many of the coding tools contributing to the coding efficiency improvements in H.266/VVC can be accelerated using SIMD instructions. With support for larger coding tree block (CTB) sizes, different coding tree structures including quad-tree (QT) & multi-type tree (MTT), more transform types and sizes, additional loop filters, more intra prediction angles and many other new coding tools, H.266/VVC standard is significantly more complex than previous ones and more efforts are required for fully accelerating H.266/VVC using SIMD. This will only become more complex with the requirement to support higher bit depths and more chroma formats. In this section, we investigate the impact of SIMD acceleration on H.266/VVC processing blocks including IQ/IT, intra prediction, inter prediction, ALF/CCALF, DBLK, SAO, LMCS filters. All SIMD implementations have been developed with recent x86 AVX2 extensions [8].

**Table 1** Performance speedup factor with SIMD optimization

| VVC Processing Block | 16bit SIMD/ non-SIMD | 8bit SIMD/ non-SIMD | 8bit SIMD/ 16bit SIMD |
|---|---|---|---|
| Intra Predict | 1.30 | 1.33 | 1.03 |
| Inter Predict | 3.66 | 4.03 | 1.09 |
| IQ/IT | 5.29 | 5.09 | 1.03 |
| LMCS | 2.17 | 2.03 | 1.10 |
| DBLK | 1.27 | 1.22 | 1.01 |
| SAO | 4.30 | 7.52 | 1.65 |
| CCALF | 9.16 | 18.10 | 1.89 |
| ALF | 12.12 | 14.82 | 1.09 |

Table 1 lists the average performance speedup factor for the H.266/VVC processing blocks with SIMD optimization in our decoder. We also demonstrate that a specialized SIMD implementation for 8-bit input bit depth can get better SIMD efficiency than 16-bit version. For some processing blocks, like CCALF and SAO, 32-bit intermediate precision is required for higher than 8-bit input, while for 8-bit input, a 16-bit intermediate precision is sufficient, so a SIMD version for 8-bit input can achieve higher register utilization and thus improves performance significantly.

### 3.2 Bit Depth Templatization

A general H.266/VVC decoder needs to be able to decode video streams with up to 10-bit as required by H.266/VVC Main10 profile. Since different bit depth would require different memory layout and structure, it becomes a critical software architecture design decision in order to support it efficiently. In H.266/VVC VTM reference software, regardless of the input bit depth, it always uses internal 16-bit memory structure which would make the software design relatively simple, but at the cost of performance loss and memory footprint increase when the input is 8bit.

Our software decoder has been designed in a more elegantly way by using C++ class and function template with bit depth as the templatized parameter, and at run time, either a 16-bit or 8-bit decoding path would be selected based on the input bit depth. This allows the decoder to easily support different input bit depth without performance penalties. Besides SIMD optimization, a major speedup can be achieved by using multiple threads to run decoding operations in parallel. Similar to H.265/HEVC, H.266/VVC also adopts several parallelization tools including slices and tiles. However, the major drawback of this level of parallelism is that the number of slices/tiles is determined by the encoder. The decoder has no control of the data dependency in the bitstream. Additionally, using multiple slices/tiles may reduce the coding efficiency.

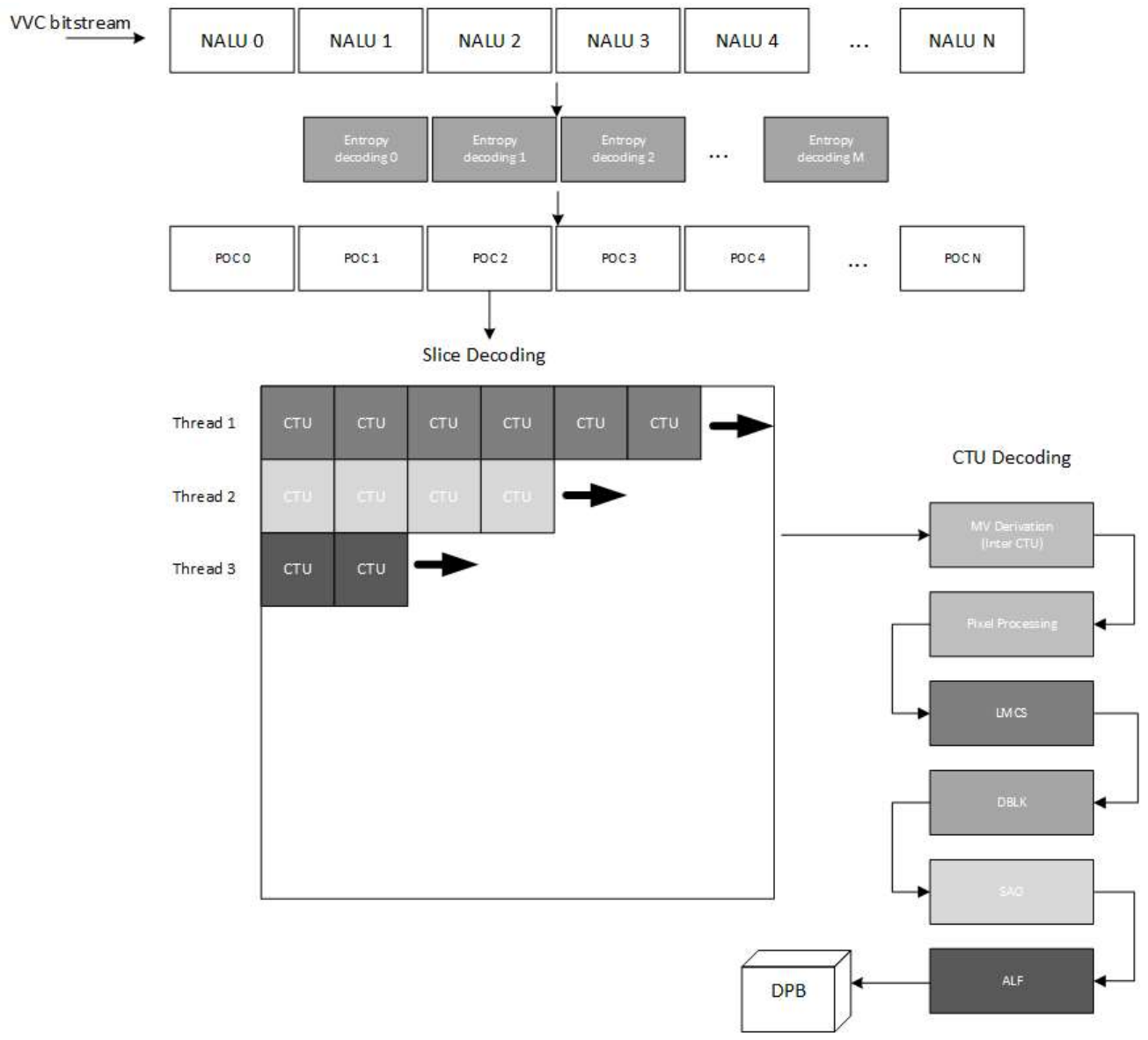

**Fig. 4** Decoder Parallelization Architecture

Our decoder implementation has employed multiple level parallelization strategies as shown in Fig.4, including picture level, CTU level, task level and sub-CTU level. Picture level parallelism consists of decoding multiple picture nal units (NALU) at the same time when the pictures are intra coded or they satisfy the motion compensation dependencies. In addition to this, entropy decoding and slice decoding of different pictures can also be processed in parallel. CTU level parallelism processes the CTUs located at different CTU rows in the wavefront pattern at the same time without breaking the coding dependencies. For the task parallelization, multiple tasks of the decoder pipeline can be executed at the same time, for example loop filtering and pixel processing (IQ/IT, intra/inter prediction and reconstruction) for different CTUs can be processed in parallel. And the sub-CTU level parallelization allows us to process multiple CUs within one CTU at the same time, for example, all the inter CUs can be processed in parallel once the motion vectors are derived for them.

For the thread management and scheduling, a thread pool with a FIFO job queue is created based on the number of cores available in the system and it consists of a group of worker threads that wait for work on the job queue. The generation of work in the job queue is dynamic and asynchronous. When all the dependencies for a new decoding job are resolved, this job is pushed to the job queue. Then any free thread from the job queue can take this job and process it.

## 4. EXPERIMENTAL RESULTS

In this section, results of our optimized software decoder described in Section 3 are presented and discussed. First, the experimental setup is described in Subsection 4.1. In Subsection 4.2, performance of our optimized H.266/VVC software decoder and the VTM reference software decoder is compared for both single thread and multithreaded executions. Unlike in H.265/HEVC where screen content coding tools such as Intra Block Copy [9] were included in the SCC extension [10], they are included in H.265/VVC version 1 due to the risen market needs. Subsection 4.3 demonstrates the performance of our software decoder with SCC tools enabled.

### 4.1 Experimental Setup

Our decoder has support for multiple operating systems including Linux, Microsoft Windows and Apple Mac OSX, the experiments presented in this paper has been performed under Linux (Ubuntu 18.04).

**Table 2** Experimental System Setup

| System | Software |
|---|---|
| Processor: Intel i9-10940X | VTM Encoder: VTM-10 |
| u-architecture: Cascade Lake | VTM Decoder: VTM-10 |
| Cores: 14 | Compiler: GCC7.5 |
| Threads/Core: 2 | C++: C++ 11 |
| Frequency: 3.3GHz | OS: Ubuntu 18.04 |
| TurboBoost: Disabled | |
| CPUFreq Governor: Performance | |

The system used for performance benchmark is a Dell Precision 5820 tower workstation equipped with an Intel Core i9-10940x processor. The processor is based on the Cascade Lake microarchitecture running at 3.30GHz CPU frequency with turbo boost disabled and CPUFreq governor set to performance mode. The processor has 14 physical cores and is able to run up to 28 threads with hyperthreading feature. Details information regarding hardware/software setup are listed in Table 2.

The test bitstreams in this section are generated by VTM-10 reference software [11], using the H.266/VVC common test conditions (CTC) [5] and H.266/VVC SCC CE test conditions [12], where UHD and HD test sequences are selected.

### 4.2 Performance Comparison of Our Decoder and VTM

Table 3 shows decoding runtimes in frames per second(fps) results between our optimized H.266/VVC software decoder running in single-thread mode and the VTM decoder with RA configuration from UHD and 1080p test sequences, respectively. It can be observed that the speedup is rather constant over different QP values, and the speedup achieved with 8-bit input is about 12% higher than 16-bit input due to our specialized 8-bit SIMD optimization as discussed in Section 3.1.

Table 4 shows the fps results when multi-threading is employed in our decoder, for all tested UHD and 1080p sequences, four, eight and sixteen worker threads are used. It can be seen from the result, that 4k 60fps real-time decoding can be achieved with 16 threads.

**Table 3** Single Thread Performance in Frames per Second (fps**)**

| Sequences | QP | VTM (fps) | Ours (fps) |
|---|---|---|---|
| UHD | 22 | 2.04 | 5.43 |
| | 27 | 2.72 | 7.06 |
| | 32 | 3.17 | 8.09 |
| | 37 | 3.64 | 9.43 |
| UHD 8-bit | 22 | 2.03 | 6.04 |
| | 27 | 2.66 | 7.71 |
| | 32 | 3.10 | 8.84 |
| | 37 | 3.53 | 10.26 |
| 1080p | 22 | 7.63 | 20.84 |
| | 27 | 11.45 | 30.12 |
| | 32 | 13.94 | 36.34 |
| | 37 | 16.24 | 41.82 |
| 1080p 8-bit | 22 | 7.64 | 23.23 |
| | 27 | 10.99 | 32.91 |
| | 32 | 13.36 | 39.25 |
| | 37 | 15.68 | 46.46 |

**Table 4** Multi-Thread Performance in Frames per Second (fps)

| Sequences | QP | VTM (fps) | Ours (fps) | | |
|---|---|---|---|---|---|
| | | | Thread 4 | Thread 8 | Thread 16 |
| UHD | 22 | 2.04 | 18.80 | 34.81 | 53.83 |
| | 27 | 2.72 | 23.92 | 44.45 | 65.46 |
| | 32 | 3.17 | 27.14 | 50.83 | 73.18 |
| | 37 | 3.64 | 30.73 | 57.80 | 80.74 |
| UHD 8-bit | 22 | 2.03 | 20.98 | 39.19 | 61.56 |
| | 27 | 2.66 | 26.28 | 49.80 | 76.19 |
| | 32 | 3.10 | 29.64 | 55.95 | 85.33 |
| | 37 | 3.53 | 34.00 | 63.96 | 95.24 |
| 1080p | 22 | 7.63 | 69.96 | 125.92 | 189.34 |
| | 27 | 11.45 | 96.99 | 175.44 | 247.22 |
| | 32 | 13.94 | 113.58 | 205.36 | 290.58 |
| | 37 | 16.24 | 128.23 | 233.78 | 331.31 |
| 1080p 8-bit | 22 | 7.64 | 76.10 | 138.86 | 213.61 |
| | 27 | 10.99 | 101.81 | 189.32 | 279.31 |
| | 32 | 13.36 | 117.96 | 221.90 | 333.46 |
| | 37 | 15.68 | 137.28 | 259.91 | 381.56 |

### 4.3 SCC Decoding Performance

Intra block copy (IBC) is a coding tool that can significantly improve the coding efficiency of screen content materials. In addition to that, block differential pulse coding modulation (BDPCM) is also included in VVC as part of the SCC coding tool set. Table 5 shows the fps results of SCC decoding performance for 1080p screen content sequences, with SCC tools enabled. At the same QP, it was observed that decoder can achieve higher performance (fps) while decoding SCC sequences compared with non-SCC ones at the same resolution because of their relatively lower complexities.

**Table 5** SCC decoding Performance in Frames per Second (fps)

| QP | VTM (fps) | Ours (fps) | | | |
|---|---|---|---|---|---|
| | | Thread 1 | Thread 4 | Thread 8 | Thread 16 |
| 22 | 16.30 | 49.64 | 141.76 | 235.65 | 311.50 |
| 27 | 17.54 | 54.90 | 154.88 | 260.06 | 336.38 |
| 32 | 18.89 | 59.42 | 166.93 | 281.84 | 361.33 |
| 37 | 20.42 | 62.69 | 178.32 | 299.86 | 383.84 |

## 5. VIDEO PLAYER

To demonstrate the proposed optimized decoder capability, we implemented a H.266/VVC video player application based on the open source VLC media player framework [13] with our decoder library as an ffmpeg plugin. The player application can decode the H.266/VVC elementary streams and render it to the output device in real-time, and we have made the player application as an open source.

## 6. FUTURE WORK AND CONCLUSION

This paper presented a highly optimized, real-time H.266/VVC software decoder. Its complexity and performance comparison with the VTM reference decoder are also demonstrated. It has been shown that 4k real-time 60fps decoding can be achieved with SIMD optimization and multi-threading task/data level parallelization. It is expected that in future developments, further speedup could be obtained by using most recent AVX512 SIMD extension and more exploration on sub-CTU level parallelization.